\documentclass[showpacs,aps,prb,groupedaddress,twocolumn,saj]{revtex4}

\usepackage{graphicx}
\usepackage{latexsym}
\usepackage[dvips]{color}
\usepackage{amsmath}
\usepackage{amssymb}

\begin{document}
\unitlength 1 cm
\newcommand{\up}{\uparrow}
\newcommand{\down}{\downarrow}
\newcommand{\nn}{\nonumber}
\newcommand{\be}{\begin{equation}}
\newcommand{\ee}{\end{equation}}
\newcommand{\bearr}{\begin{eqnarray}}
\newcommand{\eearr}{\end{eqnarray}}
\newcommand{\cdag}{c^\dagger}
\newcommand{\esite}{\bigcirc}
\newcommand{\usite}{\bigcirc\!\!\!\!\!\!\up}
\newcommand{\dsite}{\bigcirc\!\!\!\!\!\!\down}
\newcommand{\udsite}{\bigcirc\!\!\!\!\!\!\!\up\!\down}
\newcommand{\ket }[1] {|#1\rangle}
\newcommand{\bra}[1] {\langle{#1}|}

\definecolor{red}{rgb}{1.0,0.0,0.0}
\definecolor{green}{rgb}{0.0,1.0,0.0}
\definecolor{blue}{rgb}{0.0,0.0,1.0}
\definecolor{gold}{rgb}{0.85,0.66,0.0}

\title{Introduction to Hubbard Model and Exact Diagonalization}
\author{S. Akbar Jafari }
\affiliation{
Department of Physics, Isfahan University of Technology, Isfahan 84156, Iran}


\begin{abstract}
Hubbard model is an important model in theory of strongly correlated electron 
systems. In this contribution we introduce this model along with numerically
exact method of diagonalization of the model.
\end{abstract}


\maketitle  
\pagebreak
\section{Introduction}
Hubbard model\cite{Hubbard} and its variants constitute an important 
research topic in theoretical condensed matter physics, particularly 
in the context of strongly correlated electron systems.
Most of the many-body techniques commonly used in condensed matter
physics can be learnt in this context. Also there are some theoretical
tools and concepts which apply to this model only.

   There are already some monographs~\cite{Rasetti,Montorsi},
which can be used by experts, along with some text books~\cite{Fazekas,Giamarchi,Essler}
which can be consulted for further details of the various methods.
In this contribution our aim is to provide a smooth introduction
to the model and exact diagonalization technique used in dealing with Hubbard model.
This work is based on the set of lectures given by the author in 
the first IUT school\footnote{Isfahan University of Technology, June 2007} 
on strongly correlated electron systems.

   Analytical methods of solving the Hubbard model are all approximate,
except in 1D, where the so called {\bf Bethe ansatz} provides an exact
solution~\cite{LiebWu}. On the other hand there are exact numerical techniques,
which are however, either computer time expensive or memory expensive.
Therefore one is limited to rather small cluster sizes.

   A popular method to solve Hubbard model and many other models
in condensed matter physics is {\bf exact diagonalization} (ED) of the models
for small clusters which we will study at length in this set of 
lectures. We encourage the reader to implement the method presented
here in a simple \verb|fortran| program. 
In numerically exact diagonalization method one gets the
'exact' results at a high price, namely limitation to very small
cluster sizes (about 18 sites for Hubbard model at half filling),
which is essentially due to limited amount of computer RAM one
can typically have. If one accepts some error bars in numerical
results (which can however be systematically improved), then
a family of the so called {\bf Monte Carlo} methods are methods
of choice~\cite{qmc-rmp}. These methods are essentially exact. The accuracy
of the results depends on how much computer time one would 
likes to spend. In this sense, these family of methods are
time expensive, while ED is  memory expensive.

One of the important methods to deal with almost any model is 
Mean Field Theory (MFT). MFT
ignores quantum fluctuations; hence becoming less accurate in
lower spatial dimensions. Despite this, mean field treatment reveals
the wealth of various condensed matter phases can emerge from a 
simple Hubbard model~\cite{Fazekas}. In MFT one ignores both spatial and temporal
fluctuations. It is possible to retain the temporal fluctuations
by performing a full fledged quantum dynamics of the problem.
This is the subject of the so called 
{\bf 'dynamical' mean field theory}~(DMFT)~\cite{DMFT}.

   There are also a class of approximate analytic methods
known as {\bf auxiliary particle} or {\bf slave particle}
methods, which are invented to deal with the large $U$ limit
of the Hubbard model~\cite{Lavagna}. This type of techniques
are related to the so called {\bf Gutzwiller projection} 
which are devised to  obtain approximate ground
state of the Hubbard model at half filling. Generalization of
this method to deal with excited states is also there 
in the market~\cite{Muthukumar}.

  In the limit of large $U$, the charge fluctuations in the Hubbard
model are frozen and only the spin of electrons can fluctuate. Thereby
reducing the physics of the Hubbard model to spin physics described
by the so called t-J Model~\cite{Fazekas}.

\section{Hubbard model}
\subsection{Non-interacting electrons}
\label{noninteracting.sec}
The Hamiltonian of a system of non-interacting fermions
on a lattice of $L$ sites labeled by $i,j$, etc can be 
represented in second quantization by
$$
   H_0=\sum_{ij} t_{ij} c^\dagger_i c_j 
   \label{H0.eqn}  
$$
where $c^\dagger_j$($c_j$)  creates (annihilates) a fermion in a 
single-particle orbital $\phi_j$ localized at site $j$. 
In condensed matter applications one can assume $\phi_j$'s to be
Wannier wave functions (Fourier transform of Bloch orbitals).
Fermionic operators satisfy the {\em anti-commutation} relations,
\be
   \{\cdag_i,c_j\}=\delta_{ij},~~~\mbox{others}=0.
\ee
The coefficients $t_{ij}$ characterize the single-particle matrix
elements
\bearr
    t_{ij} &=&\langle \phi_i|\left(-\frac{\hbar^2\nabla^2}{2m}+\hat v\right)|\phi_j\rangle\nn\\
    &=& \int dx\phi_i(x)\left(-\frac{\hbar^2\nabla^2}{2m}+v(x)\right)\phi_j(x)
\eearr
For many practical purposes it suffices to assume that $t_{ij}$ is none-zero, only 
when $i,j$ are nearest neighbors in which case it is usually denoted by $-t$,
so that the Hamiltonian written in manifestly hermitian format becomes
\be
   H_0=-t\sum_{\langle i,j\rangle} c^\dagger_i c_j + c^\dagger_j c_i
   \label{H0t.eqn} 
\ee
Assuming the periodic boundary conditions (PBC), the Hamiltonian of the
system will be invariant under translation. The irreducible representations
of the translation group (due to abelian structure
of the group), are one-dimensional (i.e. numbers of type $e^{i\theta}$).
Hence the one-particle (or non-interacting=free) Hamiltonian (\ref{H0t.eqn})
can be diagonalized by a Fourier transformation ($\theta\leftrightarrow kj$)
\be
    c^\dagger_k =\frac{1}{\sqrt L} \sum_j e^{ikj} c^\dagger_j,
\ee
Then the Hamiltonian in $|\phi_k\rangle$ basis becomes
\be
    H_0=\sum_k \varepsilon_k c^\dagger_k c_k,~~~\varepsilon_k=-2t\cos k,
\ee
where $\varepsilon_k$ determines a cosine dispersion and represents a
band structure with $L$ allowed $k$ values in the first Brillouin zone.
Of course a simple cosine may not be a good approximation
of the realistic band structure of solids (spaghetti plots).
To mimic the realistic band structures, one can add further
neighbors' hoppings which generate higher harmonics of the simple
cosine band.
\begin{figure}[ht] 
   \centering 
   \includegraphics[height=0.33\textwidth,angle=0]{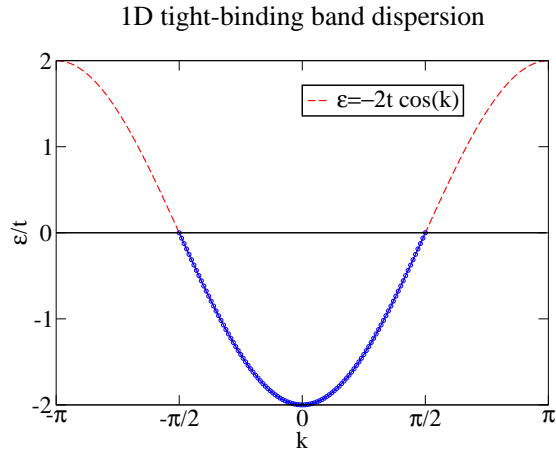}
   \caption{A tight-binding band dispersion in 1D with electron-like
   Fermi surface.}
   \label{cos.fig} 
\end{figure}

If the number $N$ of the electrons is equal to number $L$ of the sites,
then each allowed $k$ state can be occupied by two $\up$ and $\down$
spins. Hence the ground state of $H_0$ is constructed by filling the lower half 
($\varepsilon_k<0$) of the band dispersion of Fig.~\ref{cos.fig} which
is denoted by circles on the figure. 
Since half of the band is filled,  
the $N=L$ situation is called {\em half-filling}\footnote{
Note that if fermions to be filled in the band states where spin-less,
the band would be completely filled.}. 
This state is known as a Fermi sea state, usually denoted by $|{\rm FS}\rangle$. 
Second quantized representation of this state is:
\be
   |{\rm FS}\rangle = \prod_{k<k_F} \cdag_{k\up} \cdag_{k\down}|0\rangle,
\ee
where $|0\rangle$ is vacuum state (empty lattice) and $k_F$ is the largest
occupied $k$ value ($\pi/2$ here), known as Fermi wave-vector. This expression in first
quantized notation corresponds to an slater determinant. So the ground state
of a non-interacting Hamiltonian $H_0$ is characterized by a single slater
determinant.

   This state is an eigen state of the Hamiltonian 
\be
   H_0 |{\rm FS}\rangle = E_0 |{\rm FS}\rangle,
\ee
where $E_0$ is the total energy given by 
\be
   E_0=\sum_{|k|<\pi/2,\sigma} \varepsilon_k n_{k\sigma}
   =\int_{-\pi/2}^{\pi/2} \frac{dk}{2\pi} 2\times (-2t\cos k),
\ee
where the single-particle energy summation is carried out over the occupied
$k$ states of Fig.~\ref{cos.fig} shown by circles.
Obviously the depicted state has the lowest energy, since the $k$ states
have been filled such that lower energy single-particle states are filled
first.  In this spirit and excited state is create as follows: leave on 
state $k$ with $|k|<\pi/2$ empty, promoting its electron to another state
$k+q$ such that $|k+q|>\pi/2$. This excited with a hole left behind in 
state $k$ and an electron created in state $k+q$ with, say $\up$ spin
is called a 'particle-hole' excitation:
\be
   H_0|\psi_n\rangle = H_0  \cdag_{k+q\up} c_{k\up} |{\rm FS}\rangle 
   =\varepsilon^{\rm ph}_{k,q} |\psi_n\rangle
\ee
which is again an eigen state of $H_0$ with energy 
$\varepsilon^{\rm ph}_{k}(q)=\varepsilon_{k+q}-\varepsilon_k$.
This excitation carries a center of mass momentum $q$.
The above discussion can be straightforwardly generalized to higher dimensions.

Evidently the half-filled state $|{\rm FS}\rangle$ characterizes a metal,
as the energy of the particle-hole excitation can be made arbitrarily small.
The above band picture always gives a metal for an odd (in this case one) number 
of electrons per unit cell. However, as we will argue in the following,
there might be situations in which this simple prediction of the
band theory fails drastically.

\subsection{Electron-electron interaction}
When we have only $H_0$ (Eq.~\ref{H0t.eqn}) in the Hamiltonian, then
the minimization of energy is achieved by filling the $k$ states {\em independently}
with electrons of opposite spins. Such $k$ space picture in real space translates
to equal probabilities $p=1/4$ for four possible occupation of a single site:
\be
   \esite~~~
   \usite~~~
   \dsite~~~
   \udsite
   \label{sites.eqn}
\ee

   The most general form of interaction in second quantization representation
which can be added to $H_0$ is of the form
\be
   V=\frac{1}{2}\sum_{\mu\nu\alpha\beta} 
   V_{\mu\nu\beta\alpha}\cdag_{\mu} \cdag_{\nu} c_{\alpha} c_{\beta}
\ee
where $\alpha=\{\alpha,i\}$ is a collective name for the site index $i$,
and spin index $\sigma$ the two-particle matrix elements is given by
\be
   V_{\mu\nu\beta\alpha}=\int dxdx' \psi_{\mu}^*(x)\psi_{\nu}^*(x')
   V(|x-x'|) \psi_{\beta}(x) \psi_{\alpha}(x')
\ee
usually in metals with appreciable density of states $D(\epsilon_F)$ at
the Fermi level, the Coulomb potential $V(|x-x'|=r)$ is screened and
obtains the form
\be
   V(r)=\frac{e^{-rk_{\rm TF}}}{r},
\ee
where $k_{\rm TF}^{-1}$ is the so called screening length.
For the $d$ electron systems where the overlap between the atomic 
wave functions is small (smaller $t$), one has narrower band which 
is synonymous to larger DOS at the Fermi level the screening
length is usually on the scale of the Bohr radius $a^0_B$.
Therefore the most important term among all possible $\mu\nu\beta\alpha$
matrix elements is when all the indices correspond to the same site $j$.
In such case, the Pauli principle forces $\mu=\alpha=\up$ and $\beta=\nu=\down$.
Denoting the corresponding matrix element with $-2U$ one gets
for the screened interaction
\be
   V=-U\sum_j \cdag_{j\up}\cdag_{j\down}c_{j\up} c_{j\down}
   =U\sum_{j} n_{j\up} n_{j\down}.
\ee

   If we add this term to $H_0$, we obtain the celebrated 
{\em Hubbard model}
\be
   H=-t\sum_{\langle i,j\rangle\sigma}
   \left(\cdag_{i\sigma} c_{j\sigma}+\cdag_{j\sigma}c_{i\sigma}\right)
   +U\sum_j n_{j\up}n_{j\down}.
   \label{Hubbard.eqn}
\ee

   For the Hamiltonian in $U\to\infty$ limit 
the doubly occupied configuration of a single site in Eq.~(\ref{sites.eqn}) 
is going to cost a large energy $Un_{j\up}n_{j\down}=U(1)(1)$ for each doubly
occupied site. Therefore presence of such term violates the 
equi-probability of four possible states of a single-site, thereby
inducing some kind of correlation by minimizing the double occupancy.
Therefore at half-filling, and for large $U$ the ground state charge
distribution adjust itself to avoid doubly occupied sites as much
as possible; i.e. each site is occupied by a single electron, and that
moreover, large value of $U$ makes charge fluctuations around $n_{j\sigma}=1$
configuration very expensive. Therefore charge fluctuations are frozen
and one has an insulator known as {\em Mott insulator}.

   For finite values of $U$, the two terms of the Hamiltonian (\ref{Hubbard.eqn}) 
compete with each other. The kinetic energy term (corresponding to $U=0$) tends to 
delocalize 
electrons by putting individual electrons in Bloch states. 
This limit known as band limit always describes a metal. The $U$ term on the 
other hand increases the cost of charge fluctuation, leading to an insulator
in the opposite limit $U\to\infty$. Therefore there must be a critical value 
$U_c$ of the order of band width $W=2zt$ (where $z$ is the coordination number),
beyond which one has an insulator. This (first oder at zero temperature) phase
transition is known as Mott metal-insulator transition (MIT).
The only technique which can handle this model at arbitrary values of
$U$ and for arbitrary filling is ED, which will be described in next section.

\section{Exact diagonalization}
The easiest way to describe the essence of this method is by the example of 
a two-site Hubbard model. This toy model consists in two sites labeled 
$0,1$. In this case the Hubbard model written explicitly (in units in which
$t=1$) reads:
\bearr
   H&=&H_t+H_U \nn\\
   &=& -\left(\cdag_{0\up}c_{1\up}+\cdag_{1\up}c_{0\up}
   +\cdag_{0\down}c_{1\down}+\cdag_{1\down}c_{0\down}\right) \nn\\
   &&+U\left(n_{0\up}n_{0\down} + n_{1\up}n_{1\down}\right)
   \label{Hubbard2site.eqn}
\eearr
where sites are labeled as follows:
$$
   \underbrace{\square}_1 
   \underbrace{\square}_0
$$
where the site index increases from right to left.

\subsection{Organizing the Hilbert space}
To organize the Fock space for this Hamiltonian, one first notes
that the number operator $N=\sum_{j\sigma}n_{j\sigma}$ commutes with the 
Hamiltonian. Therefore one can consider only Hilbert space corresponding 
to a fixed value of $N$. Let us consider $N=2$ for this toy model which
corresponds to half-filling condition. The next question arises with regard
to the total spin of the electrons: whether they are $\up\up$, $\down\down$
or $\up\down$? First two cases represent triplet state, while the last one
corresponds to $S=0$ (more precisely $(\up\down-\down\up)$ is a singlet). 
Formally it can be checked that the total 
$S_z=1/2\sum_j\left(n_{j\up}-n_{j\down}\right)$ also commutes with the
Hamiltonian and hence a conserved quantity. Therefore there would be no
matrix element of the Hamiltonian connecting sections of the Hamiltonian
with different values of $S_z$. The structure of the Hamiltonian will be
block diagonal where each block corresponds to a fixed value of $S_z$.
To see this block-diagonal structure, we confine ourselves to $N=2$ with
both triplet and singlet spins.

  In sector with quantum numbers $N=2$ and $S_z=0$, Hilbert space 
is six dimensional with six possible basis states 
$|\phi_J\rangle$ with  $J=1\ldots 6$.
\be
\begin{array}{|c|c|c|cccc|}
  \hline  
     &   \mbox{beginner}&\mbox{expert} & \mbox{computer} & & & \\
  \hline 
  \hline J & \mbox{algebraic}&
  \mbox{picture}  & \mbox{binary} 
  & I_\down & I_\up &I \\
  \hline 1  &  \cdag_{1\up} \cdag_{0\up}|\rangle & 
    \up~~\up & |00\rangle_\down|11\rangle_\up 
  & 0 & 3&3 \\ 
  \hline 2  & \cdag_{0\down} \cdag_{0\up}|\rangle& 
   \bigcirc\up\down & |01\rangle_\down|01\rangle_\up 
  & 1 & 1&5 \\
  \hline 3  &  \cdag_{0\down}\cdag_{1\up} |\rangle&
   \up~~\down& |01\rangle_\down |10\rangle_\up 
  & 1 & 2&6 \\
  \hline 4  & \cdag_{1\down}\cdag_{0\up} |\rangle &
  \down~~\up & |10\rangle_\down |01\rangle_\up 
  & 2 & 1&9 \\
  \hline 5  &  \cdag_{1\down}\cdag_{1\up} |\rangle&
   \up\down\bigcirc& |10\rangle_\down |10\rangle_\up 
   & 2 & 2&10\\
  \hline 6  & \cdag_{0\down}\cdag_{1\down}|\rangle&
   \down~~\down & |11\rangle_\down |00\rangle_\up 
   & 3 & 0&12\\
  \hline
\end{array}
\label{basis.eqn}
\ee
First row indicates three major sets of columns:
\begin{itemize}
\item The second column labeled 
'beginner' is a way a beginner would define and work with a set
of 6 basis states $|\phi_1\rangle$ to $|\phi_6\rangle$.
\item The third column labeled 'expert' is a way an expert works
with these basis states.
\item Columns number 4 to 7 are related to the way a computer
organizes and works with the basis states.
\end{itemize}
Explanation of each columns is as follows:
\begin{enumerate}
\item In the first column there is an integer $J=1\ldots6$ which
labels the basis states in the Hilbert space. 

\item Second column shows how to obtain the configuration depicted in
third column by acting with creation operators on the vacuum $|\rangle$.
As an example look at the first basis $|\phi_1\rangle$: Physically, it describes two
$\up$ electrons in sites number $0,1$. Therefore we have two choices:
$$
   |\phi_1^a\rangle=\cdag_{1\up} \cdag_{0\up}~|\rangle,~~~\mbox{or}~~~
   |\phi_1^b\rangle=\cdag_{0\up} \cdag_{1\up}~|\rangle.
$$
These two choices by Fermionic anti commutation relations are
negative of each other: $|\phi_1^a\rangle=-|\phi_1^b\rangle$.
Then the question is:
which one is correct way of representing $|\phi_1\rangle$?
The answer is that, both of them are fine, the same way one
can describe ordinary 3 dimensional vector space with basis
$\hat e_1=\hat x,\hat e_2=\hat y,\hat e_3=\hat z$, or say,
$\hat e_1=-\hat x,\hat e_2=\hat y,\hat e_3=\hat z$.
The important point is to stick to one convention during the entire
matrix and vector manipulations. For example, to construct the above
table we choose the following convention: (i)
{\em  $\up$ spin operators sit to the right of $\down$ spins}.
(ii) {\em The order of site indices increases from right to left}. 

\item Third column is a pictorial representation of the basis vector
in real space. We will learn in the following how to work with this
intuitive representation of the basis states.

\item The fourth column represents the states as direct product 
of spin-$\down$ with spin-$\up$ state states,
where the occupations in up and down spin sectors
are represented by sequence of $0,1$ bits giving rise to a binary
representation of the basis states.

For models like, Hubbard model where $\up$ and $\down$ spins do not admix,
this separation makes a good sense.  To see this, consider the effect 
of a term like $\cdag_{i\up}c_{j\up}$ on a typical state 
  \bearr
     |\phi\rangle &=& [\cdag_{i_1\down}\ldots\cdag_{i_N\down}]~
     [\cdag_{j_1\up}\ldots \cdag_{j_M\up}]|\rangle \nn\\
     &=& [\cdag_{i_1\down}\ldots\cdag_{i_N\down}]|\rangle_\down
     ~ [\cdag_{j_1\up}\ldots \cdag_{j_M\up}]|\rangle_\up\nn\\
     &=:& |\phi_\down\rangle |\phi_\up\rangle\nn,
  \eearr
which can be written as
\bearr
   &&\cdag_{i\up} c_{j\up}|\phi_\down\rangle |\phi_\up\rangle =
   \cdag_{i\up} c_{j\up}[\cdag_{i_1\down}\ldots\cdag_{i_N\down}]
   [\cdag_{j_1\up}\ldots \cdag_{j_M\up}]|\rangle \nn\\
   &&= [\cdag_{i_1\down}\ldots\cdag_{i_N\down}]~
   (-1)^N (-1)^N \cdag_{i\up} c_{j\up}
   [\cdag_{j_1\up}\ldots \cdag_{j_M\up}]|\rangle\nn\\
   &&= |\phi_\down\rangle~ \cdag_{i\up}c_{j\up} |\phi_\up\rangle.
   \label{uphop.eqn}
\eearr
Here two $(-1)^N$ factors arise from moving each of the $\cdag_\up$
operators through a length $N$ sequence of $\cdag_\down$ operators. Therefore, 
{\em when operating in spin-$\up$ sector, we need not worry about spin-$\down$ 
configuration and vice versa}.

\item The last three column indicate the way a computer understands and
stores these basis state. There is only one integer $I$ stored on computer
which fully represents the occupation pattern of the $\up$ and $\down$ 
electrons when transformed to binary representation of fourth column in 
Eq.~(\ref{basis.eqn}). Given $I$, one can obviously
find out $I_\up$ and $I_\down$ through the relation 
\be
   I=2^LI_\down+I_\up
\ee 
and vice versa. First and last columns of (\ref{basis.eqn}) 
do actually tabulates an array $I=T(J)$ which for  any given $J$, 
returns the corresponding $I$. The value of $I$ fully specifies the state.
One therefore only to extract the bits of $I$ and find a way to work with
bits of integer $I$.
\end{enumerate}

There remains only a final note on how we have labeled the states.
We have labeled the configuration represented by $I=9$ as fourth
($J=4$) basis vector, etc. We could have labeled them in any order,
so that $I=9$ would have corresponded to, e.g. first ($J=1$) basis vector.
The above convention which was suggested by Lin and coworkers\cite{Lin},
has the advantage that the table T can be searched given a value for $I$
in a fast way in order to find out corresponding $J$ value. The essential idea
of this convention is the following~\cite{Lin}: For any sector you are interested in, just 
choose the labels $J$ in such a way that when the above table T is constructed,
the $I$ values are ascending function of $J$.

\subsection{Acting with operators on the basis states}
If the action of an operator on a complete basis set is known, then the
operator is completely specified. Eq.~(\ref{Hubbard2site.eqn}) has
two type of terms: $H_U$ and $H_t$. When $H_U$ acts on a basis
states, it gives non zero contribution for each site $j$ in which
both $n_{j\up}$ and $n_{j\down}$ are $1$. Therefore the effect of
$H_U$ on any basis state gives the same state multiplied by
the number of doubly occupied sites $\times U$.  Hence the effect of 
$H_U$ on $|\phi_i\rangle,~i=1,3,4,6$ is zero and 
on $|\phi_2\rangle$ and $|\phi_5\rangle$ is just $U$. The $H_U$
part is diagonal in occupation number representation:
\be
   H_U\doteq U \left(\begin{array}{cccccc}
   0 & 0 & 0 & 0 & 0 & 0 \\
   0 & 1 & 0 & 0 & 0 & 0 \\
   0 & 0 & 0 & 0 & 0 & 0 \\
   0 & 0 & 0 & 0 & 0 & 0 \\
   0 & 0 & 0 & 0 & 1 & 0 \\
   0 & 0 & 0 & 0 & 0 & 0 
   \end{array}\right)
   \label{HUmat.eqn}
\ee

Now let us concentrate on the effect of $H_t$ on our basis states.
Consider, e.g. the sate $|\phi_2\rangle$. $H_t$ in Eq.~(\ref{Hubbard2site.eqn})
has $4$ terms. But only $2$ of them give non zero contribution when they
act on $|\phi_2\rangle$: The one which allows an $\up$ spin to hop from
site $0$ to $1$, i.e. $\cdag_{1\up}c_{0\up}$, and another one which 
allows the $\down$ spin at site $0$ to jump to site $1$, i.e.
$\cdag_{1\down}c_{0\down}$.
\bearr
   &&H_t |\phi_0\rangle = -t\left(\cdag_{1\up}c_{0\up}
   +\cdag_{1\down}c_{0\down}\right)\cdag_{0\down} \cdag_{0\up}|\rangle\nn\\
   &&=-t\left(\cdag_{1\up}\overbrace{c_{0\up}}^{\rightarrow}\cdag_{0\down} \cdag_{0\up}
   +\cdag_{1\down}c_{0\down}\cdag_{0\down} \overbrace{\cdag_{0\up}}^{\leftarrow\leftarrow}\right)|\rangle\nn\\
   &&= -t\left(-\cdag_{1\up}\cdag_{0\down}c_{0\up} \cdag_{0\up}|\rangle
   +\cdag_{1\down}\cdag_{0\up}c_{0\down}\cdag_{0\down}|\rangle \right)\nn\\
   &&= -t\left(-\cdag_{1\up}\cdag_{0\down}|\rangle
   +\cdag_{1\down}\cdag_{0\up}|\rangle \right)\nn
\eearr
where in the third line we have used the Fermionic anti-commutation 
relations with associated minus signs needed for each exchange of
Fermionic operators with different indices.
Also in the last step we have used the fact that
$c_{j\sigma}\cdag_{j\sigma}|\rangle=(1-n_{j\sigma})|\rangle=|\rangle$, as
the vacuum state $|\rangle$ contains no particles. Rearranging the 
Fermionic operators to comply with our convention we get
\be
   H_t|\phi_2\rangle = -t \left(+|\phi_3\rangle+|\phi_4\rangle \right).
   \label{Htphi2.eqn}
\ee
This way of working with commutations and algebra is not a convenient 
for putting on computers. As we already showed in Eq.~(\ref{uphop.eqn}),
the up and down spins hop separately. Therefore, when an spin $\up$ electron
hops from site say $j$ to site $i$, one only needs to count a $(-1)$ factor
for each $\up$ spin electron over which it passes. Since when $c_{j\up}$
has to moved through a chain of $\cdag_{j'\up}$ operators with $j'\ne j$,
until it reaches $\cdag_{j\up}$, there they form a $(1-n_{j\up})$ operator
which commutes with all other remaining $\cdag_{j''\up}$ operators to reach
the vacuum $|\rangle$ where it produces $|\rangle$ itself.

   With this argument in mind, one can most conveniently work with
the pictorial representation of states: The effect of $H_t$ on 
$|\phi_2\rangle$ is to either move an $\up$ spin to site $1$, or
to move a $\down$ spin to site $1$. In the former case one gets 
$(-1)^0|\phi_3\rangle$ where the exponent $0$ is because the $\up$
spin at site $0$ passes through no other $\up$ spin when it hops to
site $1$. Similarly the later case gives $|\phi_4\rangle$. According
to the Hamiltonian (\ref{Hubbard2site.eqn}), each of 
these processes happens with an amplitude $-t$  and thus one can pictorially
see that 
$$
   H_t|\phi_2\rangle = -t \left(+|\phi_3\rangle+|\phi_4\rangle \right).
$$

With this in mind it is almost trivial to check that
\bearr
   && H_t |\phi_1\rangle = 0\nn\\
   && H_t |\phi_2\rangle = -t\left(|\phi_3 \rangle + |\phi_4 \rangle\right) \nn\\
   && H_t |\phi_3\rangle = -t\left(|\phi_2 \rangle + |\phi_5 \rangle\right) \nn\\
   && H_t |\phi_4\rangle = -t\left(|\phi_2 \rangle + |\phi_5 \rangle\right) \nn\\
   && H_t |\phi_5\rangle = -t\left(|\phi_3 \rangle + |\phi_4 \rangle\right) \nn\\
   && H_t |\phi_6\rangle = 0
   \label{Htphi.eqn}
\eearr
Therefore in this basis the hopping term has the following
matrix representation:
\be
   H_t\doteq -t \left(\begin{array}{cccccc}
   0 & 0 & 0 & 0 & 0 & 0 \\
   0 & 0 & 1 & 1 & 0 & 0 \\
   0 & 1 & 0 & 0 & 1 & 0 \\
   0 & 1 & 0 & 0 & 1 & 0 \\
   0 & 0 & 1 & 1 & 0 & 0 \\
   0 & 0 & 0 & 0 & 0 & 0 
   \end{array}\right)
   \label{Htmat.eqn}
\ee
Therefore the matrix representation of the entire Hamiltonian in this
basis becomes
\be
   H\doteq  \left(\begin{array}{c|cccc|c}
   0 & 0 & 0 & 0 & 0 & 0 \\ \hline
   0 & U & -t & -t & 0 & 0 \\
   0 & -t & 0 & 0 & -t & 0 \\
   0 & -t & 0 & 0 & -t & 0 \\
   0 & 0 & -t & -t & U & 0 \\ \hline
   0 & 0 & 0 & 0 & 0 & 0 
   \end{array}\right)
   \label{Hmat.eqn}
\ee
Clearly the above block diagonal structure has a parallel with the following 
form for the $S_z$ matrix:
\be   
    S_z\doteq\left(\begin{array}{c|cccc|c}
    +1 &   &   &   &   &    \\ \hline
    & 0  &   &   &   &    \\
    &   & 0  &   &   &    \\
    &   &   & 0  &   &    \\
    &   &   &   & 0  &    \\ \hline
    &   &   &   &   & -1   
   \end{array}\right)
\ee
In other words state $|\phi_1\rangle$ has a quantum number
$S_z=+1$, while $|\phi_6\rangle$ has $S_z=-1$. The set of states 
$|\phi_2\rangle\ldots|\phi_5\rangle$ belong to $S_z=0$ sector.
If we had confined ourselves to $S_z=0$ sector, we would have
obtained a four dimensional Hilbert space in which the Hamiltonian
could be represented by the $4\times 4$ block of Eq.~(\ref{Hmat.eqn}).

  The method described after Eq.~(\ref{Htphi2.eqn}) has the advantage
that for large matrices, precisely the same steps can be taken with a 
simple computer code that implements the logic described above.
When coding the procedure we have the binary
representation in column 4 of Eq.~(\ref{basis.eqn}) in mind for the occupation
pattern of $\up$ and $\down$ electrons. But there is only one integer $I$ (7th 
column) stored on computer the binary representation of which precisely
corresponds to column $4$. Extracting  $I_\up$ and $I_\down$ from
a given $I$ in \verb|fortran| is as simple as 
\begin{verbatim}
   I_up=mod(I,2**L);  I_dn=I/2**L
\end{verbatim}
Most of the standard high level programming
languages such as \verb|c++| and \verb|fortran 90| have appropriate
intrinsic functions for bitwise operations on integer which are 
able to access and examine or change individual bits of a given integer
$p$ (can be $I_\up$ or $I_\down$).
Therefore an integer \verb|I| completely specifies the occupation pattern
in each spin sector. For example the Intel Fortran compiler has following
commands for bitwise manipulation of an integer \verb|p|
\be
\begin{array}{|l|l|}
   \hline
   \verb|IBSET(p,b)|  & \cdag_b\\
   \verb|IBCLR(p,b)|  & c_b\\
   \verb|BTEST(p,b)| & n_b\\
   \hline
\end{array}
\ee
For more details please consult the Intel Fortran language manual~\cite{intelmanual}.
Also more implementation notes can be found in Lin et.al.~\cite{Lin}.

\subsection{Diagonalization of the matrix: Space symmetries}
So far we have made use of the symmetries of the {\em Hamiltonian} itself
to reduce the dimension of the matrix to be diagonalized.
For example conservation of $S_z$ (equivalent to $[S_z,H]=0$) reduces
the $6\times 6$ matrix to two $1\times 1$ (trivial) and one $4\times 4$
matrix. The commutation relation between e.g. $S_z$ and $H$ is 
quite general, and independent of the geometry of the lattice
in use. 

   Now we would like to make use of the spatial symmetries
in order to diagonalize the $4\times 4$ matrix in $S_z=0$ sector manually.
For a two site problem composed of sites $0,1$, there is a mirror reflection
operator $M$ which has the following action on real lattice:
\be
   M(0)=1,~~M(1)=0
\ee
Corresponding to operation $M$ which is a member of spatial symmetry
group of the underlying lattice, there is an operator ${\cal M}$ in 
Hilbert space which acts on the state vectors. For example consider
the effect of $\cal M$ on state $|\phi_3\rangle$ which reads
\bearr
  &&{\cal M} \cdag_{0\down} \cdag_{1\up}|0\rangle
  =\cdag_{M(0)\down} \cdag_{M(1)\up}|0\rangle
  =\cdag_{1\down} \cdag_{0\up}|0\rangle,
\eearr
or ${\cal M}|\phi_3\rangle=|\phi_4\rangle$. Similarly,
${\cal M}|\phi_4\rangle=|\phi_3\rangle$, etc. 
which is summarized as follows
\bearr
   {\cal M}|\phi_1 \rangle = -|\phi_1 \rangle\nn\\
   {\cal M}|\phi_2 \rangle = |\phi_5 \rangle\nn\\
   {\cal M}|\phi_3 \rangle = |\phi_4 \rangle\nn\\
   {\cal M}|\phi_4 \rangle = |\phi_3 \rangle\nn\\
   {\cal M}|\phi_5 \rangle = |\phi_2 \rangle\nn\\
   {\cal M}|\phi_6 \rangle = -|\phi_6 \rangle\nn
\eearr
The first and last lines of the above equation indicate that
states $|\phi_j\rangle$ with $j=1,6$ are eigen states
of ${\cal M}$ (parity operator) with eigen values $\pm 1$.
Line numbers 3,4 however, indicate that states $|\phi_3\rangle$
and $|\phi_4\rangle$ do not have definite parity. Instead
a new combination 
$|\phi_\pm\rangle=\frac{1}{\sqrt 2}(|\phi_3\rangle\pm|\phi_4\rangle)$,
has a definite parity: ${\cal M}|\phi_\pm\rangle=\pm|\phi_\pm\rangle$.
Similarly from $|\phi_2\rangle$ and $|\phi_5\rangle$, symmetric and
antisymmetric combinations will have definite parity.

   The eigenvalues $\pm$ of parity operator $\cal M$ partition 
the Hilbert space into two pieces which belong either to $+1$
eigen value:
\bearr
   &&{\cal M}\left(|\phi_3\rangle+|\phi_4\rangle\right)= 
   +\left(|\phi_3\rangle+|\phi_4\rangle\right)\nn\\ 
   &&{\cal M}\left(|\phi_2\rangle+|\phi_5\rangle\right)= 
   +\left(|\phi_2\rangle+|\phi_5\rangle\right),
   \label{evenparity.eqn}
\eearr
or to $-1$ eigen value:
\bearr
   &&{\cal M}|\phi_1 \rangle = -|\phi_1 \rangle\nn\\
   &&{\cal M}|\phi_6 \rangle = -|\phi_6 \rangle\nn\\
   &&{\cal M}\left(|\phi_3\rangle-|\phi_4\rangle\right)= 
   -\left(|\phi_3\rangle-|\phi_4\rangle\right)\nn\\ 
   &&{\cal M}\left(|\phi_2\rangle-|\phi_5\rangle\right)= 
   -\left(|\phi_2\rangle-|\phi_5\rangle\right)
   \label{oddparity.eqn}
\eearr
If we knew this when finding out the effect of $H_t$ in Eq.~(\ref{Htphi.eqn}),
we would have acted on the following states:
\bearr
   &&|\psi_1\rangle=|\phi_1\rangle\nn\\
   &&|\psi_2\rangle=\left(|\phi_2\rangle+|\phi_5\rangle\right)/\sqrt 2\nn\\
   &&|\psi_3\rangle=\left(|\phi_3\rangle+|\phi_4\rangle\right)/\sqrt 2\nn\\
   &&|\psi_4\rangle=\left(|\phi_3\rangle-|\phi_4\rangle\right)/\sqrt 2\nn\\
   &&|\psi_5\rangle=\left(|\phi_2\rangle-|\phi_5\rangle\right)/\sqrt 2\nn\\
   &&|\psi_6\rangle=|\phi_6\rangle 
   \label{symmetrybasis.eqn}
\eearr
Eq.~(\ref{symmetrybasis.eqn}) defines the so called, symmetry
adopted basis in which the action of, say, $H_t$ is enormously 
simplified:
\bearr
   && H_t |\psi_1\rangle = 0\nn\\
   && H_t |\psi_2\rangle = -2t|\psi_3 \rangle \nn\\
   && H_t |\psi_3\rangle = -2t|\psi_2 \rangle \nn\\
   && H_t |\psi_4\rangle = 0\nn\\ 
   && H_t |\psi_5\rangle = 0\nn\\ 
   && H_t |\psi_6\rangle = 0
   \label{Htpsi.eqn}
\eearr
Similarly the only none zero matrix elements of $H_U$ are
given by:
\be
   H_U|\psi_2\rangle = U|\psi_2\rangle,
   ~~~H_U|\psi_5\rangle = U|\psi_5\rangle 
\ee

How do we generate a symmetry adopted basis? There is a very powerful
theorem in group representation theory
which in case of one dimensional representations is easy to implement
and reads~\cite{Slater}:
\be
   \psi^{(p)} = \sum_R \Gamma_p(R)^* R\phi 
\ee
In this equation $\phi$ is an arbitrary state to begin with.
$R$ is an element of the symmetry group which in this case
and be either ${\cal I}$ or ${\cal M}$. $\Gamma_p$ is the
irreducible representation of the group which for group 
$\{{\cal I},{\cal M}\}$ are numbers $\pm 1$ for even and odd 
parity. Reader can easily check that feeding $\phi=|\phi_i\rangle$,
with $i=1,\ldots,6$ is going to generate the symmetry adopted basis
Eq.~(\ref{symmetrybasis.eqn}). Similar technique can be used
to generate states with definite $\vec k$ values when dealing
with problems of translational invariance.

In symmetry adopted basis total Hamiltonian in $S_z=\pm 1$ 
sectors ($\{|\psi_1\rangle, |\psi_6\rangle\}$ sub-space)
remains diagonal with eigen values equal to $0$, as it was.
In $S_z=0$ sector it will become:
\be
   H\doteq  \left(\begin{array}{cc|cc}
   U & -2t & 0 & 0 \\
   -2t & 0 & 0 & 0 \\ \hline
   0 & 0 & 0 & 0 \\
   0 & 0 & 0 & U \\ 
   \end{array}\right)
   \label{Hmatpsi.eqn}
\ee
In this sector as well, there are one eigen value equal 
to $0$ which corresponding to eigen state
$|\psi_4\rangle$. Note that $|\psi_4\rangle$
is the $S_z=0$ component of a triplet. Therefore the triplet state
lies at zero energy:
\be
   E_t=0,~~~ \mbox{for triplet state }(S_z=+1,0,-1)
   \label{Et.eqn}
\ee
The fact that $\{|\phi_1\rangle,|\phi_5\rangle,|\phi_6\rangle\}$
form a triplet is consistent with having odd spatial parity
in Eq.~(\ref{oddparity.eqn}).

The eigen value corresponding to $|\psi_5\rangle$ is $U$ which
is always positive and above the triplet $E_t=0$.

To find out the remaining two eigen values in the 
$\{|\psi_2\rangle, |\psi_3\rangle\}$ sector, we note that first
of all, this sector has even spatial parity, and hence
is spin singlet. Hamiltonian is
$U/2{\cal I}+U/2\sigma_z-2t\sigma_x$, where $\sigma_x,\sigma_z$ 
are  Pauli matrices, and  ${\cal I}$ is the unit $2\times 2$ matrix,
so that the eigenvalues at spin singlet sector become
\be
  E_s^{\pm}=U/2\pm\sqrt{(U/2)^2+4t^2}
\ee
Therefore the ground state is a singlet with energy
$E_s^-$, while the first excited state is a triplet
with $E_t=0$. Second excited state is $|\psi_5\rangle$,
with energy $U$, and final excited state has energy $E_s^+$.

\section{Strong correlations and spin physics}
As we already saw in previous section, the ground state 
of the two site Hubbard model is a singlet with energy
$E_s^-=U/2-\sqrt{U^2/4+4t^2}$, The ground state wave function
\be
   |E_s^-\rangle = 
   4|\psi_2\rangle+\left(U+\sqrt{U^2+16}\right)|\psi_3\rangle
\ee
in large $U$ limit is dominated by 
$|\psi_3\rangle\sim|\phi_3\rangle+|\phi_4\rangle$ in which
there is no doubly occupied configuration, and hence 
charge fluctuations are suppressed.

 Since the first excited state
is at $E_t=0$. The splitting between these two states for
large $U\gg t$ is:
\be
   -J=E_s-E_t=U/2-\sqrt{U^2/4+4t^2}\approx -\frac{4t^2}{U}
\ee
Therefore the singlet state is slightly below ($-4t^2/U$) the
triplet state. This indicates that in large $U$ limit, the
low-energy physics of Hubbard model is given by spin fluctuations
which are anti ferromagnetic (singlet has lower energy).
This observation in a two site Hubbard model is indeed very
general and it can be shown using a unitary transformation that
the Hubbard model at large $U$ limit can be mapped into the
so called t-J model, where there are AF spin fluctuations
along with hoppings restricted to subspace with no double
occupancy~\cite{Fazekas}.

\end{document}